\documentclass[aps,prd,twocolumn,showpacs,superscriptaddress,showkeys,nofootinbib]{revtex4-1}

\usepackage{amssymb}
\usepackage{amsmath}
\usepackage{graphicx}
\usepackage{amsfonts}
\usepackage{color}
\usepackage{xspace}
\usepackage{ulem}
\usepackage{mathtools}
\usepackage{hhline}
\usepackage[caption=false]{subfig}
\usepackage{footnote}
\usepackage{hyperref}
\usepackage{adjustbox}

\newcommand{\AddrCoimbra}{Univ Coimbra, Faculdade de Ci\^encias e Tecnologias da Universidade de Coimbra and CFisUC, Rua Larga, 3004-516 Coimbra, Portugal}
\begin{document}

\title{Superradiant Darwinism: survival of the lightest axion}

\author{Daniel Neves}
\email{nevesdanielf@gmail.com}
\author{Jo\~{a}o G.~Rosa } \email{jgrosa@uc.pt} \affiliation{\AddrCoimbra}   

\date{\today}

\begin{abstract}
We study the dynamical evolution of superradiant instabilities of rotating black holes for multiple axion fields with comparable masses, motivated by string theory constructions, which typically exhibit a large number of light axions, with a broad range of masses. We show, in particular, that even though superradiant clouds for the heavier axion species grow faster, they are eventually reabsorbed by the black hole as the latter amplifies the lighter axion field(s), analogously to the dynamics of different species competing for the same resources in an ecosystem. We also incorporate in our study the effects of accretion and gravitational wave emission. We further demonstrate that the existence of multiple axion species with comparable masses may have a substantial impact on the stochastic gravitational wave background produced by axion clouds around black hole binary merger remnants, which could be probed with planned detectors.
\end{abstract}


\maketitle

\section{Introduction}

Rotational black hole (BH) superradiance (SR) has been widely studied in the recent literature, particularly as a probe of ultra-light axion fields \cite{Arvanitaki:2009fg, Arvanitaki:2010sy, Rosa:2012uz, Witek:2012tr, Brito:2014wla, Arvanitaki:2016qwi,  Brito:2017wnc, Rosa:2017ury, Baumann:2018vus,   Ikeda:2018nhb, Berti:2019wnn, Baumann:2019eav, Cardoso:2018tly, Edwards:2019tzf,Mehta:2020kwu,Blas:2020nbs,Yuan:2021ebu, Tong:2022bbl, Siemonsen:2022ivj} or other low mass particles \cite{Rosa:2011my, Pani:2012bp, Brito:2013wya,Baryakhtar:2017ngi,Caputo:2021efm}. The growing evidence for the existence of stellar mass and supermassive BHs, namely from gravitational wave detection \cite{LIGOScientific:2016aoc, LIGOScientific:2018mvr, LIGOScientific:2020ibl, KAGRA:2021vkt} and radio imaging of galactic centers \cite{EventHorizonTelescope:2019dse, EventHorizonTelescope:2022wkp}, further motivates exploring their potential as fundamental physics laboratories.

The basic idea is that Kerr BHs may become unstable if they spin too fast, shedding their excess angular momentum into a cloud of bosonic particles  gravitationally bound to the BH. This is reminiscent of Press and Teukolsky's ``BH bomb'' idea \cite{Press:1972zz, Cardoso:2004nk}, whereby co-rotating wave modes of frequency $\omega<m\Omega_H$, where $\Omega_H$ denotes the BH horizon's angular velocity and $m$ the wave's azimuthal angular momentum number, are amplified upon scattering off the BH and then reflected by a surrounding mirror multiple times, thus creating an instability. 

For massive fields, the corresponding wave modes can be trapped in quasi-bound states in the gravitational potential well created by the BH \cite{Damour:1976kh, Zouros:1979iw, Detweiler:1980uk, Furuhashi:2004jk, Cardoso:2005vk, Dolan:2007mj, Rosa:2009ei, Dolan:2012yt, East:2017ovw, East:2017mrj, Dolan:2018dqv}. This confinement effectively plays the same role as Press and Teukolsky's reflecting mirror, leading to a continuous amplification of the field and therefore to an exponential growth of the corresponding number of bound particles. Since the wave frequency coincides in this case with the field's mass $\mu$ (in natural units) up a small binding energy, the superradiance condition above translates, for the leading co-rotating modes with $m=1$, into a lower bound on the BH dimensionless spin parameter, above which it becomes superradiantly unstable:
\begin{equation}
\tilde{a} >{4\alpha\over 1+4\alpha^2}~,    
\end{equation}
where $\alpha=\mu M/ M_P^2$ is the dimensionless mass coupling, with $M$ denoting the BH mass and $M_P$ the Planck mass. This yields an upper bound on the field's mass, since superradiance is only possible for $\alpha< m/2$ taking into account that $\tilde{a}\leq 1$ \footnote{Assuming that naked singularities are unphysical.}. For the leading modes this yields:
\begin{equation}
\mu< 6.6\times10^{-11}\left({M_\odot\over M}\right) \ \mathrm{eV}. 
\end{equation}
Hence, known astrophysical BHs can only become superradiantly unstable if there exist in Nature bosonic fields\footnote{Pauli blocking prevents fermionic superradiance \cite{Unruh:1973bda}.} with masses well below those of currently known elementary particles. The most well-motivated candidates are axion-like particles, which are naturally light scalars since their mass is protected by an underlying perturbative shift symmetry that is only broken by exponentially small non-perturbative effects. 

It has been argued, in fact, that realistic string theory scenarios harbour $\mathcal{O}(10^2-10^3)$ axions \cite{Arvanitaki:2009fg}, arising from dimensional reduction of the extra-dimensional Ramond-Ramond and Neveu-Schwarz $p$-form gauge fields, and which inherit perturbative shift symmetries from the underlying 10-dimensional gauge invariance, making the axions massless at tree level. While certain moduli stabilization mechanisms (e.g.~that preserve supersymmetry) could make these axions heavy already at tree level, this would spoil the Peccei-Quinn solution to the strong CP problem \cite{Peccei:1977hh, Peccei:1977ur} by changing the QCD axion \cite{Weinberg:1977ma, Wilczek:1977pj} potential. It follows that string compactifications including at least one light QCD axion should in fact yield several hundred axions, with exponentially small masses resulting exclusively from non-perturbative effects. 

On the one hand, due to its non-perturbative nature, the scale of a typical string axion potential is exponentially suppressed relative to the theory's UV scale\footnote{This typically high energy scale depends on the mechanism generating the axion potential.}, $\Lambda^4 = M_{UV}^4e^{-S}$, where $S$ denotes the corresponding euclidean instanton action. String axion decay constants, on the other hand, are inversely proportional to this action, $f_a\sim M_P/S$, so that the corresponding mass is
\begin{equation}
\mu\sim {\Lambda^2\over f_a} \sim {M_{UV}^2\over M_P} Se^{-S/2}~.
\end{equation}
The instanton action $S$ is typically related to the area of the cycle in the compact extra-dimensions associated with each axion (in units of the fundamental string scale) up to $\mathcal{O}(1)$ factors. For non-QCD instantons, $S\gtrsim 200$ or otherwise these could spoil the Peccei-Quinn mechanism by modifying the QCD axion potential. This means that, while all axion decay constants should lie close to the GUT scale $M_{GUT}\sim 10^{16}$ GeV if the compact manifold is not too anisotropic, their masses should be distributed over several decades in magnitude.

As already observed in the seminal string axiverse article \cite{Arvanitaki:2009fg}, this means that we may expect several axions per mass decade, including some with masses differing by $\mathcal{O}(1)$ factors. As we will show in this work, this may have a significant impact on the dynamics of axion superradiant instabilities, essentially because, as we will revise in the next section, the e-folding time characterizing the growth of scalar superradiant clouds is approximately proportional to $\mu^9$ (for the dominant superradiant mode) for a given BH mass and spin. This implies that axion fields with masses differing by over an order of magnitude (within the superradiant regime) grow on widely different timescales. For instance, a BH with 10$M_\odot$ and $\tilde{a}=0.7$, will grow a superradiant axion cloud (which takes $\mathcal{O}(100)$ e-folds) in about a century for $\mu\simeq 10^{-12}$ eV, while this would take longer than the age of the Universe for $\mu\simeq 10^{-13}$ eV. If, however, there are axions with masses differing by less than an order of magnitude, a single BH may potentially grow multiple superradiant clouds since its formation until the present day.

Anticipating our main result, we will show that the superradiant production of lighter axions, albeit slower, forces the reabsorption of previously formed heavier axion clouds, since they deplete the BH spin below the latter's superradiance threshold. This effectively works as a natural selection mechanism within the string axiverse, with lighter axions being the ``fittest'' ones for a spinning BH, and thus the ones that may potentially survive until the present day. We then discuss several effects that could influence the dynamics of multiple axion clouds and observational prospects for gravitational wave emission.

This work is organized as follows. In the next section we review the dynamics of superradiant instabilities for a single massive scalar field, deriving a new approximate analytical result for the time evolution of the particle number. We then discuss the dynamics with two scalar fields of comparable masses in section III, generalizing to an arbitrary number of fields and also including the effects of an accretion disk and gravitational wave emission. In section IV we study the potential detectability of gravitational waves emitted by multiple superradiant axion clouds, focusing in particular on the stochastic background resulting from clouds formed around remnants of binary BH mergers. We summarize our main results and conclusions in section V.
We use natural units $\hbar=c=1$ throughout our discussion.


\section{Single axion BH superradiance}

In the vicinity of a rotating BH, a massive scalar field $\Phi$ (namely an axion-like field) can form Hydrogen-like quasi-bound states described by the quantum numbers $(n, l, m)$ (see \cite{Brito:2015oca} for a review). These are analogous to bound states in a Coulomb potential, since this is the asymptotic form of the BH's gravitational potential, but with the dimensionless mass coupling $\alpha=\mu M/M_P^2$ playing the role of the fine structure constant. In the non-relativistic regime where $\alpha\ll1$ in which we will be most interested in, the real part of the quasi-bound state frequencies is then given approximately by:
\begin{equation}
    \omega_{R,n} \simeq \mu-{\alpha^2\over 2n^2}\mu~,
\end{equation}
The quasi-bound state frequencies are, however, complex as a result of the field modes penetrating the BH horizon as ingoing waves, with the field being damped/amplified depending on whether the imaginary part of the frequency is negative/positive, since $\Phi\propto e^{-i\omega t}$. In the non-relativistic regime, the bound state occupation numbers decrease/grow at an approximate rate given by:
\begin{equation}
\label{eq: SRrate}
\begin{aligned}
\Gamma_{nlm} & =-\left(\frac{l !}{(2 l+1) !(2 l) !}\right)^2 \frac{(l+n) !}{(n-l-1) !} \frac{4^{2 l+2}}{n^{2 l+4}}  \\
& \times \prod_{k=1}^l\left(k^2+16\left(\frac{M (\omega_R-m\Omega_H)}{\tau}\right)^2\right) \\ & \times  \left(\frac{r_{+}-r_{-}}{r_{+}+r_{-}}\right)^{2 l+1} {\alpha^{4 l+5}\over \tau}\left(\omega_R-m\Omega_H\right)~,
\end{aligned}
\end{equation}
where $\Omega_H = \tilde{a}/(2r_+)$ is the angular velocity of the BH event horizon with dimensionless spin parameter $\tilde{a}= JM_P^2/M^2$, $r_\pm= M\left(1\pm \sqrt{1-\tilde{a}^2}\right)$ are the radial coordinates of the inner (Cauchy) and outer (event) horizons and $\tau=(r_+-r_-)/r_+$. Amplification thus occurs in the superradiant regime, $\omega_R < m \Omega_H$, at the expense of extracting the BH's rotational energy and therefore decreasing its spin, until this condition is saturated when $\omega_R=m\Omega_H$. While the BH also loses mass to produce a superradiant cloud, this may be discarded in the non-relativistic regime since $\Delta M /M\simeq \alpha \tilde{a}_i\ll 1$, with $\tilde{a}_i$ denoting the initial BH spin. 

Since $\Gamma_{nlm}\propto \alpha^{4l+5}$, the leading $2p$-mode with quantum numbers $(nlm)=(211)$ grows exponentially faster than all higher-$l$ modes, so that we may neglect the latter for $\alpha\ll1$. This yields a superradiant cloud with an approximately toroidal shape with a radius $\simeq 5r_B$, where the gravitational Bohr radius is $r_B= (\alpha\mu)^{-1}\gg r_+$. The growth rate for this mode is well approximated by:
\begin{equation}
    \label{approx gamma}
    \Gamma\equiv \Gamma_{211} \simeq\frac{1}{24}(\tilde{a}-4\alpha) \alpha^8\mu~.
\end{equation}
The number of particles in the $2p$-cloud then grows as:
\begin{equation}
    \label{dNdt}
    \frac{dN}{dt} = \Gamma N~,
\end{equation}
making the BH mass and spin vary according to:
\begin{equation}
    \label{dMdt}
    \frac{dM}{dt} = -\mu \Gamma N~,
\end{equation}
and
\begin{equation}
    \label{dadt}
    \frac{d\tilde{a}}{dt} = -\left(1-2\tilde{a}\alpha\right)\left({M_P\over M}\right)^2\Gamma N~,
\end{equation}
For small $\alpha$ we may, neglecting the change in the BH mass to leading order, combine Eqs.~\eqref{dNdt} and \eqref{dadt} and integrate to yield
\begin{equation}
    \label{a n relation}
    \tilde{a} = \tilde{a}_i - \frac{M_P^2}{M^2}\left(N-N_i\right)~,
\end{equation}
where $\tilde{a}_i$ and $N_i$ denote the initial values of the BH spin and particle number.  We may then replace this relation in the approximate expression \eqref{approx gamma} for the superradiance growth rate to obtain a decoupled equation for the number of particles in the superradiant cloud:
\begin{equation}
    \frac{dN}{dt} = \frac{\mu\alpha^8}{24}\left(\tilde{a}_i -4\alpha -\frac{M_P^2}{M^2}\left(N-N_i\right)\right) N~,
\end{equation}
which we can write in the form of the {\it logistic equation}:
\begin{equation}
    \label{diff eqn: logistic}
    \frac{dN}{dt} = \gamma N\left(1 -\frac{N}{N_\text{max}}\right)~,
\end{equation}
where we have defined the parameters:
\begin{equation}
    \label{r}
    \gamma=\frac{\mu\alpha^8}{24}\left(\tilde{a}_i -4\alpha +\frac{M_P^2}{M^2}N_i\right)\simeq \frac{\mu\alpha^8}{24}\left(\tilde{a}_i -4\alpha\right)~,
\end{equation}
\begin{equation}
    \label{K}
    N_\text{max}=N_i+{M^2\over M_P^2}(\tilde{a}_i -4\alpha)\simeq {M^2\over M_P^2}(\tilde{a}_i -4\alpha)~,
\end{equation}
and in the last steps we have used that $M\gg M_P$ for the classical BHs in which we are interested.

It is well known that the logistic equation describes the growth of the population $N(t)$ in an ecosystem, with $N_\text{max}$ denoting its carrying capacity and $\gamma$ the population growth rate. This differential equation has a simple analytical solution:
\begin{equation}
    \label{eqn: logistic half}
    N(t) = \frac{N_\text{max}}{1+e^{-\gamma(t-t_{1/2})}}~,
\end{equation}
where $t_{1/2}\equiv \gamma^{-1}\log(N_\text{max}/N_i-1)$, such that $N(t_{1/2})=N_\text{max}/2$. This shows that, for $N_i\ll N_\text{max}$, the population starts growing exponentially, with $N\simeq N_ie^{\gamma t}$ for $t\lesssim t_{1/2}$, but then asymptotes to the carrying capacity $N_\text{max}$ for $t\gg t_{1/2}$. This is also true for $N_i\gg N_\text{max}$, which is therefore an attractor. The carrying capacity thus describes the maximum number of individuals, at which the population is in equilibrium with the ecosystem and above which the latter cannot provide enough food for all individuals. 

This yields a perfect analogy between superradiant axion clouds and biological ecosystems, where the BH plays the role of the environment and its spin is the food that allows the number of axions to grow. As clear in Eq.~\eqref{K}, the maximum number of axions is dictated by how fast the BH is initially spinning , i.e.~by how much food the ecosystem can provide. While initially there can be plenty of food for the population to grow exponentially fast, when it becomes close to the carrying capacity the food shortage slows the population growth down. 

In this way, even if we have only a single individual as  initial condition the number $N$ will grow until it reaches the maximum capacity $N_\text{max}$ and then stays at $N=N_\text{max}$ in equilibrium with the ecosystem. If we start with more individuals than the carrying capacity ($N_i > N_\text{max}$), some will ``die'' until the number decreases reaching $N_\text{max}$, since there is not enough food for all individuals. This means that our system, in complex dynamical systems language, has one attractive fixed point which is exactly $N=N_\text{max}$. 

\begin{figure}[htbp]
    \centering
    \includegraphics[width=0.46\textwidth, keepaspectratio]{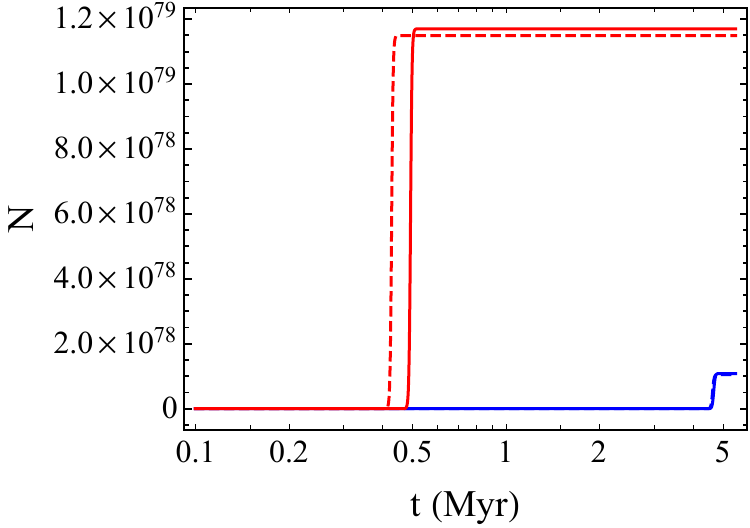}
    \caption{Numerical result for the number of axions with mass $\mu=10^{-13}$ eV in a (211)-superradiant cloud around a BH with initial mass $M_i=50M_\odot$ and spin $\tilde{a}_i=0.2$ (blue) or $\tilde{a}_i=0.7$ (red). The corresponding dashed curves yield the logistic approximation.}
    \label{fig1}
\end{figure}

To assess the validity of the approximate logistic solution, we have numerically integrated Eqs.~(\ref{dNdt})-(\ref{dadt}) starting with a single scalar particle in the (211)-bound state (e.g. from a quantum fluctuation as generically assumed). As we will see later, we are particularly interested in BHs resulting from binary mergers in the mass range detected with LIGO-Virgo-KAGRA (LVK), with a few tens of solar masses, yielding for the dimensionless mass coupling:
\begin{equation}
    \alpha \simeq 0.04 \left(\frac{M}{50M_\odot}\right)\left(\frac{\mu}{10^{-13}\ \rm eV}\right).
\end{equation}
In Figure \ref{fig1} we compare the numerical solution with the analytic logistic approximation for an axion with $\mu=10^{-13}$ eV and a BH with initial mass $M_i=50M_\odot$ and for two values of the initial spin $\tilde{a}_i=0.2$ and $0.7$.

As clear in this figure, the logistic function yields indeed a very good approximation to the growth of the superradiant cloud, albeit it somewhat underestimates the maximum number of particles. A better agreement may nevertheless be obtained if we replace the logistic parameters with $N_{\text{max}}=\Delta M/\mu$ and $\gamma=\mu\alpha_f^8N_\text{max}/M_f^2$, where $\Delta M=M_i-M_f$ and the final BH mass is given by \cite{Yuan:2021ebu}:
\begin{equation}
	\label{eq: newfinal mass}
M_f = \frac{1-\sqrt{1-16\alpha_i^2(1-\alpha_i\tilde{a}_i)^2}}{8\alpha_i^2(1-\alpha_i\tilde{a}_i)}M_i~,
\end{equation}
which can be obtained by equating $\Delta M \simeq  \mu \Delta J$ and $\Omega_H^f\simeq \mu$, and where the subscripts `i' and `f' denote the initial and final values of the parameters.


\section{Superradiance with two axions}

We now turn to the main topic of this work, which is the dynamics of BH superradiant instabilities in the presence of multiple axions with comparable masses. We start with the case of two axion fields, $\Phi_1$ and $\Phi_2$, such that $\mu_1\gtrsim \mu_2$ and hence $\alpha_1 > \alpha_2$.
 This implies that if $\Phi_1$ is in the superradiant regime, then $\Phi_2$ also obeys the SR condition. Furthermore, even when the SR condition for the heavier axion is saturated the lighter axion remains in the SR regime, since in this case $\Omega_H\simeq\mu_1> \mu_2$. In terms of the BH spin parameter,  when the SR condition is saturated for the heavier axion we have:
 \begin{equation}
    \alpha_2 < \frac{\tilde{a}_{f1}}{2\Big(1+\sqrt{1-\tilde{a}_{f1}^2}\Big)}\simeq\alpha_1.
\end{equation}
This yields $\tilde{a}_{f1}\simeq 4\alpha_1/(1+4\alpha_1^2)\simeq 4\alpha_1$ in the non-relativistic regime. Since the lighter axion remains in the SR the BH spin is reduced below $\tilde{a}_{f1}$, so that the heavier axion enters a non-superradiant regime where the corresponding SR cloud is necessarily reabsorbed by the BH. Note that reabsorption corresponds to a sign flip in the rate given approximately by Eq.~(\ref{approx gamma}).

Since the SR rates $\Gamma_i\propto \mu_j\alpha_i^8$, $j=1,2$, the lighter field is amplified much more slowly than the heavier field, and we thus expect the heavier axion SR cloud to form earlier. Notice, for instance, that a factor $2$ in the mass difference, \textit{i.e.} $\mu_1 =2\mu_2$, implies $\Gamma_1/\Gamma_2 \sim 2^9\sim 10^3$. Given that the growth of the number of particles is exponential, we may assume that the BH's spin changes, in a first stage, only due to the growth of the heavier axion SR cloud. 
This yields $\Delta \tilde{a}= \tilde{a}_i-\tilde{a}_{f1}\simeq \tilde{a}_{f1}-4\alpha_1$ neglecting the change in the BH mass as previously discussed, and we thus expect the heavier axion cloud to attain its maximum capacity $N_1^\text{max}\simeq (\tilde{a}_i-4\alpha_1)) (M/M_P)^2$ before the second field is significantly amplified.

In a second stage, once the lighter axion field $\Phi_2$ begins to grow significantly and the BH spin decreases substantially below $\tilde{a}_{f1}\simeq 4\alpha_1$, we expect the heavier axions to be quickly reabsorbed by the BH, as $\Gamma_1<0$ with $|\Gamma_1|\gg \Gamma_2$. The heavier axions (in the $l=m=1$ state) thus return their angular momentum to the BH, but this only fuels the growth of the lighter axion cloud. Thus, we expect the latter to grow as if the heavier axion cloud never formed, i.e. from a BH with spin parameter $\tilde{a}_i$, thus attaining a maximum capacity $N_2^\text{max}\simeq (\tilde{a}_i-4\alpha_2) (M/M_P)^2$. Hence, we expect:
\begin{equation}
	\label{racios}
		\frac{N^{\rm max}_2}{N^{\rm max}_1} \simeq {\tilde{a}_i-4\alpha_2\over \tilde{a}_i-4\alpha_1}
	= 1 + 4\frac{(\alpha_1-\alpha_2)}{(\tilde{a}_i- 4\alpha_1)} > 1~.
\end{equation}
To better understand the dynamics of the two-field scenario, we neglect the BH mass change as discussed in the single-field case. In this approximation we then have for the BH spin evolution:
\begin{equation}
	\frac{d\tilde{a}}{dt} =-\frac{M_P^2}{M^2}\left(\frac{dN_1}{dt} +\frac{dN_2}{dt}\right)=-\frac{M_P^2}{M^2}{dN\over dt}~,
\end{equation}
where $N=N_1+N_2$ is the total number of heavy and light axions produced in the corresponding $(211)$ bound states. This can be integrated to give simply:
\begin{equation}
	\tilde{a}= \tilde{a}_i -\frac{M_P^2}{M^2}(N-N_i)~,
\end{equation}
analogously to the single field case. Substituting this into the evolution equations for the heavy and light axion numbers, $dN_j/dt =\Gamma_j N_j$, we then have:
\begin{equation}
	\label{full system}
	\begin{split}
		{} & \frac{dN_1}{dt} = \frac{\mu_1 \alpha_1^8}{24}\left[\tilde{a}_i -4\alpha_1-\frac{M_P^2}{M^2}(N-N_i)\right]N_1
		\\ &\frac{dN_2}{dt} = \frac{\mu_2\alpha_2^8}{24}\left[\tilde{a}_i-4\alpha_2 -\frac{M_P^2}{M^2}(N-N_i)\right]N_2
	\end{split}~.
\end{equation}

We may express these equations in terms of the $\gamma_j$ and $N_j^\text{max}$ parameters for each field, defined as in Eqs.~(\ref{r}) and (\ref{K}), yielding:
\begin{eqnarray}
	\label{dn1dt simple}
	\frac{dN_1}{dt} &=& \gamma_1N_1\bigg( 1-\frac{N_1+N_2}{N^{\rm max}_1}\bigg)\nonumber\\
	\frac{dN_2}{dt} &=& \gamma_2N_2\bigg( 1-\frac{N_1+N_2}{N^{\rm max}_2}\bigg)
\end{eqnarray}
Combining these two equations gives:
\begin{equation}
	\label{helper}
\frac{N^{\rm max}_1}{\gamma_1}\frac{d\log N_1}{dt}-\frac{N^{\rm max}_2}{\gamma_2}\frac{d\log N_2}{dt} = N^{\rm max}_1-N^{\rm max}_2~,
\end{equation}
which can be integrated to obtain the relation:
\begin{equation}
	\label{n1n2 relation}
 {(N_1/N_{1,0})^{N^{\rm max}_1/\gamma_1}\over (N_2/N_{2,0})^{N^{\rm max}_2/\gamma_2}}
 = e^{(N^{\rm max}_1-N^{\rm max}_2)(t-t_i)}~.
\end{equation}
This equation thus relates the number of axions in the two SR clouds at each stage in the evolution. In principle, we could use this relation to eliminate the dependence on, \textit{e.g.}, $N_2$ in equation Eq.~($\ref{dn1dt simple}$) to obtain a non-linear non-homogeneous differential equation for $N_1$, but this would be challenging to solve analytically. 

Nevertheless, we may use this relation to determine the time at which the heavy axion SR cloud is reabsorbed and the lighter axion SR cloud forms, which we denote as the {\it transition time}. We may define this as the time at which $N_1=\epsilon N_2^{\rm{max}}$ and $N_2=(1-\epsilon)N_2^{\rm{max}}$ for some $\epsilon<1$. Setting initial conditions $N_{1,i}=N_{2,i}=1$ (at $t_i\ll t_{\text{trans}}$) we then obtain:
\begin{equation}
	\label{eq: tfinal}
	t_{\text{trans}} \simeq \frac{N^{\text{max}}_2}{\gamma_2(N^{\text{max}}_2-N^{\text{max}}_1)}\left[\log(N^{\text{max}}_2)-\left(\frac{\alpha_2}{\alpha_1}\right)^9\log\epsilon\right].
\end{equation}
Since $(\alpha_2/\alpha_1)^9\ll 1$, the precise value of $\epsilon$ does not significantly affect the transition time (which in fact shows that it is a sharp transition, as a consequence of the exponential character of the SR instability). Hence, neglecting this logarithmic dependence and also the logarithmic dependence on $\tilde{a}_i$, we find approximately:
\begin{equation}
	\label{eq: ttransition}
		 t_{\text{trans}} \simeq {} 12 \ \bigg(\frac{0.015}{\alpha_2}\bigg)^9 \Big(\frac{5\times10^{-14} \text{eV}}{\mu_2}\Big) 
		{\log(M/40M_\odot)\over r-1}\ \text{Gyr} ,
\end{equation}
where we have defined the mass ratio $r=\mu_1/\mu_2>1$ \footnote{A better agreement is found by using the final BH mass (\ref{eq: newfinal mass}).}. This shows that such transitions may be happening close to the present day ($t_0=13.8$ Gyr).

Given that the lighter axion SR clouds may grow on such long timescales, one needs to check whether excited quasi-bound states of the heavy field may grow faster than the latter. In particular, the (322) state, which is the second fastest growing mode, is still in the SR regime after the corresponding (211) cloud is fully formed, with $\tilde{a}\simeq 4\alpha_1$ as discussed above. However, it is easy to show that, at this stage:
\begin{equation}
	\label{eq: 2srmodes}
	{\Gamma_1^{(322)}\over \Gamma_2^{(211)}} \simeq 2.2\times10^{-4}\alpha_1^4 {r^{10}\over r-1}~,
\end{equation}
which is below unity for $\alpha_1< 0.2$ if the mass ratio $r=\mu_1/\mu_2<5$, i.e.~for most of the parameter space of interest to our discussion the lighter axion cloud grows faster than the excited heavy axion SR modes, so that we will discard them in our discussion henceforth\footnote{See \cite{Ficarra:2018rfu} for a dynamical study involving multiple modes.}.

\begin{figure}[h!]
    \centering
        \includegraphics[width=0.48\textwidth, keepaspectratio]{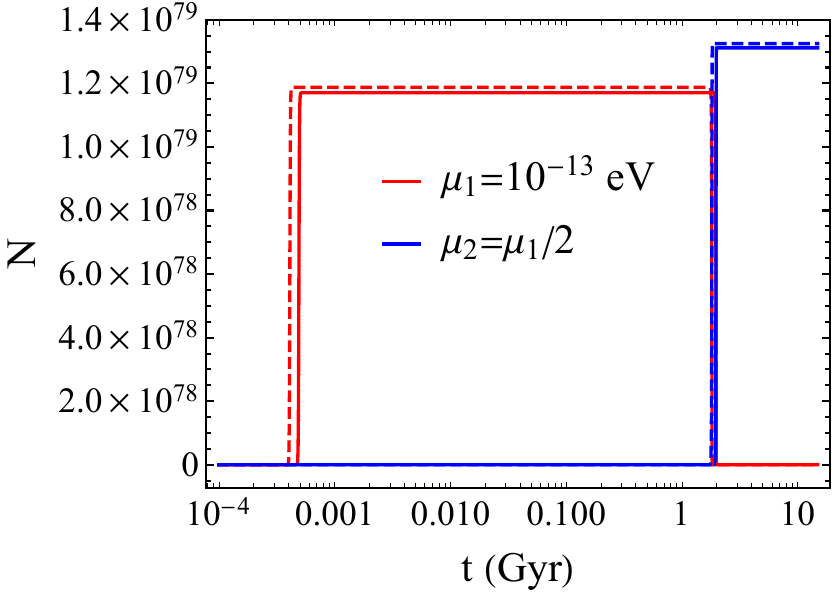}
    \caption{Numerical evolution of the number of heavy (red) and light (blue) axions in the corresponding (211)-SR clouds, for $\mu_1=10^{-13}$ eV and $\mu_2=\mu_1/2$ (blue) around a BH with $M_i=50M_\odot$ and $\tilde{a}_i=0.7$. The respective dashed curves yield the results obtained neglecting the change in the BH mass, solving Eqs. (\ref{full system}) but considering the final BH mass in (\ref{eq: newfinal mass})}. 
    \label{fig: two_field_dynamics}
\end{figure}

\begin{figure}[h!]
    \centering
    \includegraphics[width=0.474\textwidth]{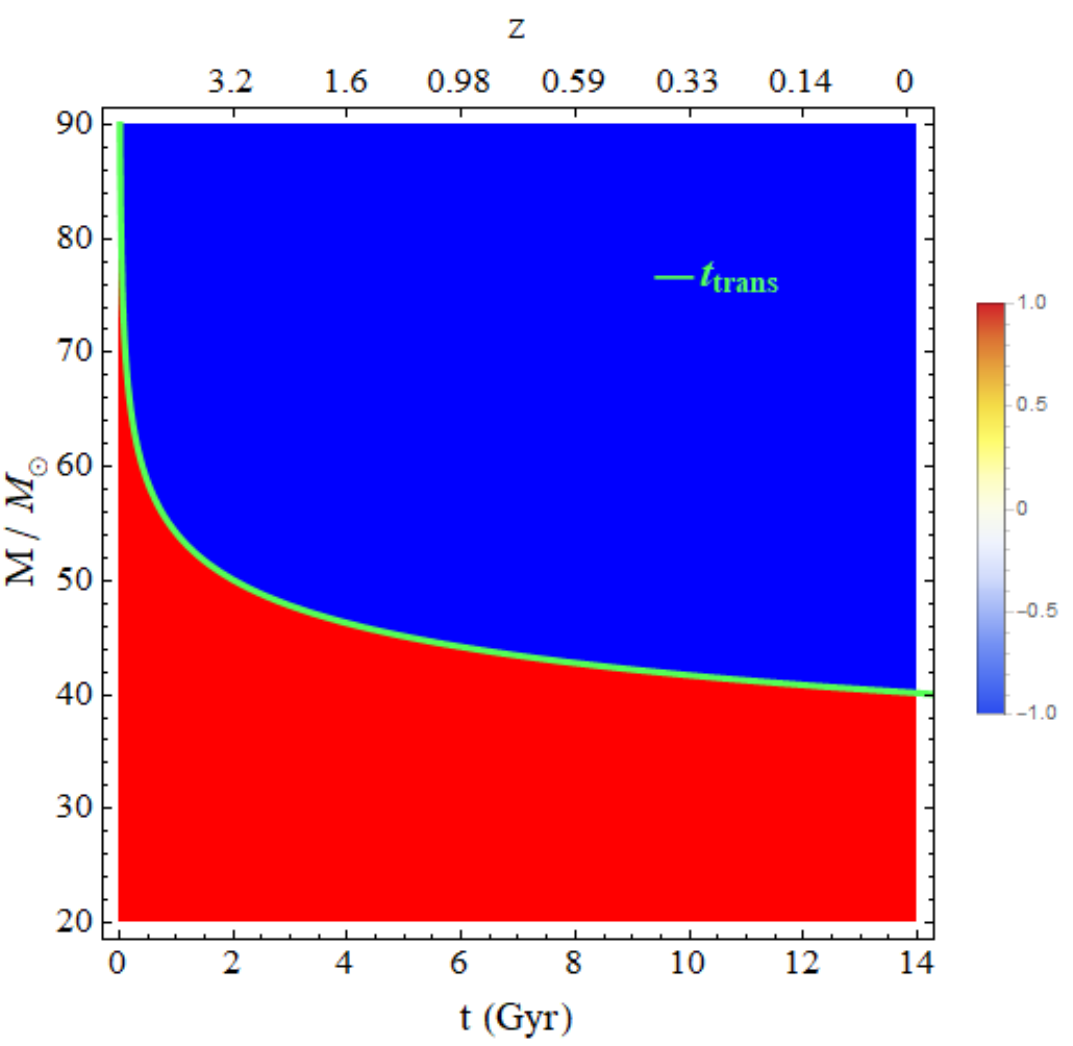}
    \caption{Results for $N_2/N_2^{\textrm{max}}-N_1/N_1^{\textrm{max}}$ as a function of time and redshift for different initial BH masses, considering $\mu_1=10^{-13}$ eV , $\mu_2=\mu_1/2$ and $\tilde{a}_i=0.7$. The red (blue) region corresponds to a fully grown heavy (light) axion cloud, while the green curve corresponds to the heavy-to-light axion cloud transition given by Eq.~(\ref{eq: tfinal}).} 
	\label{fig: heatmap}
\end{figure}

In Fig.~\ref{fig: two_field_dynamics} we show the results of numerically evolving the BH and two axions system for  $\mu_1=10^{13}$ eV, $\mu_2=\mu_1/2$, $M_i=50M_\odot$ and $\tilde{a}_i=0.7$. In Fig.~\ref{fig: heatmap}, we plot $N_2/N_2^{\textrm{max}}$-$N_1/N_1^{\textrm{max}}$, which takes values between $-1$ (fully grown heavy axion cloud) and $+1$ (fully grown light axion cloud), for different BH masses. These figures illustrate the sharpness of the transition between the heavy and light axion SR clouds, which is nearly ``instantaneous'' on cosmological timescales.

It is interesting to note that Eqs.~(\ref{dn1dt simple}) have the form of the competitive Lotka-Volterra equations describing the population dynamics of two species competing for a common resource, and which generalize the logistic equation for a single species. Hence, again in the two-field case we find an interesting analogy between BH-axion superradiance and biological ecosystems. 

In our case, both the heavy and light axion species are competing for the ``food'' provided by the BH spin. The heavier axion population grows faster and reaches a quasi-equilibrium state with the ecosystem (the BH), with the number of individuals (particles within the SR cloud) nearly reaching the maximum capacity. However, the growth of the lighter axion species, albeit slower, eventually decreases the level of food available below the minimum threshold required for the survival of the heavy axions. These essentially ``die'' and become part of the environment (i.e.~are reabsorbed by the BH), thus boosting the growth of the lighter axion population towards its maximum capacity.

Curiously, within this analogy the lighter axions require less ``food'' to grow, so in the end they naturally prevail as the dominant species. It is not difficult to see that this generalizes to an arbitrary number of axion species: the lighter species always remain in the SR regime after the heavier ones reach their maximum capacity, so they continue to grow in decreasing order of the axion mass. Thus, the SR dynamics with multiple fields exhibits a natural selection mechanism, with only the lightest axion SR cloud eventually surviving. 

Like in biological systems, the discussion so far may, however, be too simplistic to describe realistic systems, and we need in particular to consider two other effects that may potentially influence the growth of axion SR clouds: gravitational wave emission and accretion.

\subsection{Gravitational Wave emission}

Axion SR clouds, and more generally SR clouds of any real scalar field, are neither spherically symmetric nor stationary and thus possess a non-vanishing time-dependent quadrupole moment, resulting in the emission of gravitational waves (GWs). At the particle level, this can be interpreted as the result of two axions (of the same species) annihilating into a graviton (see e.g.~\cite{Arvanitaki:2010sy}). This may be relevant, in particular to the growth of the lighter axion SR cloud, given that GWs resulting from heavy axion annihilation transport angular momentum away from the system, so that a fraction of the BH spin is effectively lost instead of reabsorbed. Within our biological analogy we are effectively allowing individuals to escape the ecosystem in pairs\footnote{We leave the potential interpretation of what this could mean within a biological system to the reader.}. We could, in addition, include axion quartic self-interactions, which would also lead to dissipative effects, but in the axion mass range that we are considering these are typically too weak to be relevant even on cosmological time scales (see e.g. \cite{Baryakhtar:2020gao, Branco:2023frw}).

The rate at which a SR cloud loses energy through GW emission is given approximately by \cite{Brito:2014wla}:
\begin{equation}
	\label{eq: egw bh}
	\frac{dE_{GW}}{dt}\simeq \frac{484+9\pi^2}{23040} \left(\frac{M_P}~,{M}\right)^2M_c^2\alpha^{14},
\end{equation}
where $M_c\simeq \mu N$ is the SR cloud's mass. For a single axion species, GW emission thus modifies the evolution of the SR cloud's occupation number as:
\begin{equation}
	\label{eq: dndtgw}
	\frac{dN}{dt} = \Gamma N -\Gamma_{GW}N^2~,
\end{equation}
where $\Gamma_{GW}= \dot{E}_{GW}/(\mu N^2)\simeq\mu \dot{E}_{GW}/M_c^2$. Comparing the two terms on the right-hand side of this equation for $\tilde{a}\simeq \tilde{a}_i$, we obtain:
\begin{equation}
	{\Gamma_{GW} N^2\over \Gamma N}\simeq 0.6\alpha^6{N\over N_{\rm{max}}}~,
\end{equation}
so that GW emission does not affect the growth of the SR cloud until it is fully formed and the SR condition is saturated such that $\Gamma\rightarrow 0$ in the single field case. Hence, we may divide the dynamics in two stages: a first stage where the axion occupation number is amplified by SR and evolves according to the logistic function Eq.~(\ref{eqn: logistic half}), up to a time $t_{SR}\gtrsim t_{1/2}$ where the SR cloud (nearly) reaches its maximum capacity\footnote{In practice the maximum capacity is never reached, so we have $N(t_{SR})=(1-\epsilon)N_{\rm{max}}$ for some $\epsilon\ll 1$.}; and then a second stage driven by GW emission, for which the occupation number decreases from $N(t_{SR})\simeq N_{\rm{max}}$ according to:
\begin{equation}
\frac{dN}{dt} = -\Gamma_{GW} N^2~, 
\end{equation}
such that
\begin{equation} \label{N_GW}
N\simeq {N_{\rm{max}}\over 1+ {(t-t_{SR})\over\tau_{GW}}}~, \quad t> t_{SR}~,
\end{equation}
where $\tau_{GW}^{-1}=\Gamma_{GW}N_{\rm{max}}$ should be evaluated at the BH mass $M_f$. To give an idea of the time scales involved, we have e.g.~for $\mu=10^{-13}$ eV and a BH of $65M_\odot$ ($\alpha\simeq 0.05$) that $t_{SR}\simeq 4\times 10^{-5}$ Gyrs and $\tau_{GW}\simeq 30$ Gyrs.

In the two-axion case, the heavier axion SR cloud should evolve according to the above discussion, while GW emission from the lighter axion cloud is in general much less significant, given the large power of $\alpha$ determining the GW emission rate. The growth of the light axion cloud at $t_{\rm{trans}}$ may nevertheless be affected by the reduction of the number of heavy axions that results from GW emission. This reduction will only be significant if $\tau_{GW}\lesssim t_{\rm{trans}}$, i.e. for
\begin{equation} \label{GW_effect}
0.15 {\log(N_2^{\rm{max}})\over r-1}(\tilde{a}_i-4\alpha_1){\alpha_1^{15}\over \alpha_2^{10}}>1~,
\end{equation}
where we recall that $r=\mu_1/\mu_2>1$ and that $t_{\rm{trans}}\gg t_{SR}$. For typical values, this implies that GW emission will have a negligible impact on the two-field dynamics for $\alpha_2<	\alpha_1\lesssim  \alpha_2^{2/3}$, i.e. when the ratio between the two masses is not too large.

\begin{figure}[h]
	\centering
\includegraphics[width=0.48\textwidth,keepaspectratio]{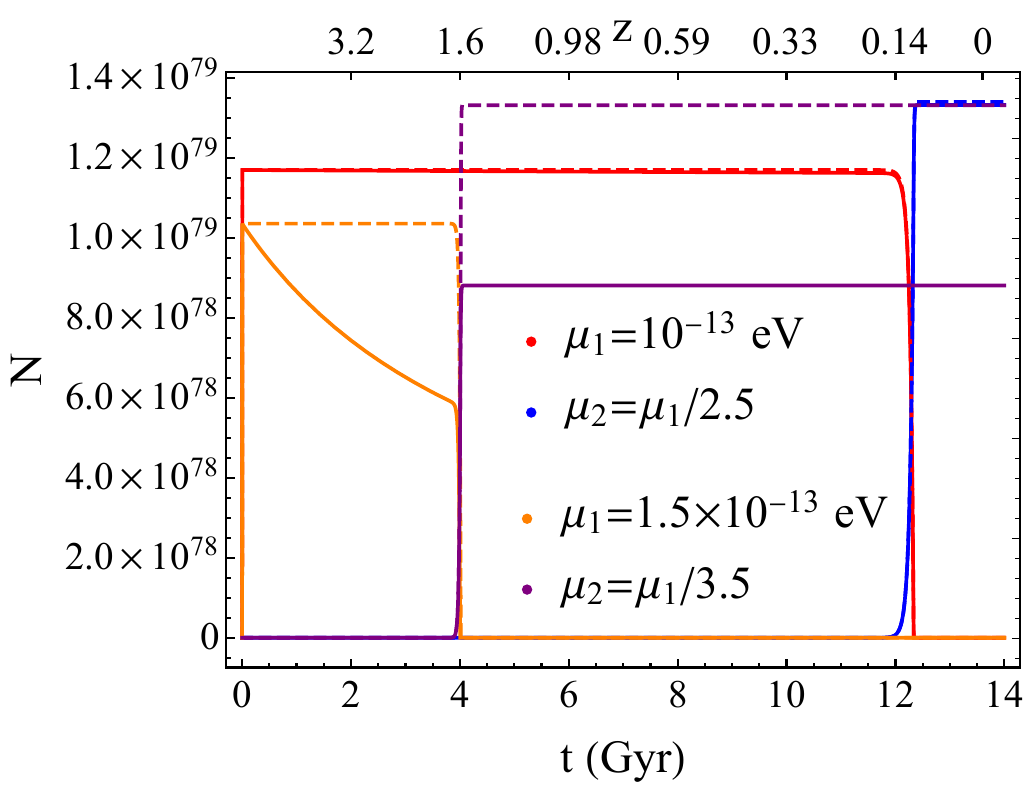}
	\caption{Numerical results for the number of heavy and light axions in the corresponding axion clouds as a function of time and redshift $z$, for $M_i=50M_\odot$ and $\tilde{a}_i=0.7$, and considering two possible sets of mass values:  $\mu_1=10^{-13}$ eV (red) and $\mu_2=\mu_1/2.5$ (blue);  $\mu_1=1.5\times10^{-13}$ eV (orange) and $\mu_2=\mu_1/3.5$ (purple). The solid (dashed) curves correspond to the evolution with (without) GW emission.}
	\label{fig: tnt}
\end{figure}

To illustrate how GW emission affects the dynamics, in Fig.~\ref{fig: tnt} we illustrate the numerical evolution of the two-axion system for a BH with $M_i=50M_\odot$ and $\tilde{a}_i=0.7$, taking two different pairs of axion masses: $\mu_1= 10^{-13}$ and $\mu_2= \mu_1/2.5$; $\mu_1= 1.5\times10^{-13}$ and $\mu_2= \mu_1/3.5$. As one can see in this figure, for the lower mass ratio GW emission does not change the dynamics significantly, whereas for the larger ratio (with smaller $\mu_2$) the heavier axion cloud is visibly depleted by axion annihilations into GWs and the lighter axion cloud cannot grow as much as in the absence of GW emission (although occupation numbers change by only an $\mathcal{O}(1)$ factor in this example). 

Taking into account that GW emission reduces $N_1$ by a factor $f_1\simeq (1+t_{\rm{trans}}/\tau_{GW})^{-1}$ relative to the maximum capacity, and the amount of angular momentum that is therefore dissipated away instead of reabsorbed by the BH during the heavy-light axion transition, we find that the lighter axion cloud is also reduced by a factor:
\begin{equation}
	f_2\simeq f_1-4(1-f_1)\frac{\alpha_1-\alpha_2}{\tilde{a}_i-4\alpha_2}.
\end{equation} 
This means that the effect of GW emission is somewhat more pronounced on the lighter axion cloud. For instance, for $f_1=0.9$ (so that 10$\%$ of the heavy axions annihilate into GWs), $\alpha_1=2\alpha_2=0.1$ and $\tilde{a}_i=0.7$, we find $f_2=0.86$, i.e.~a reduction of 14$\%$ in the number of light axions.

\begin{figure}[h!]	
	\centering
	\includegraphics[width=0.48
\textwidth]{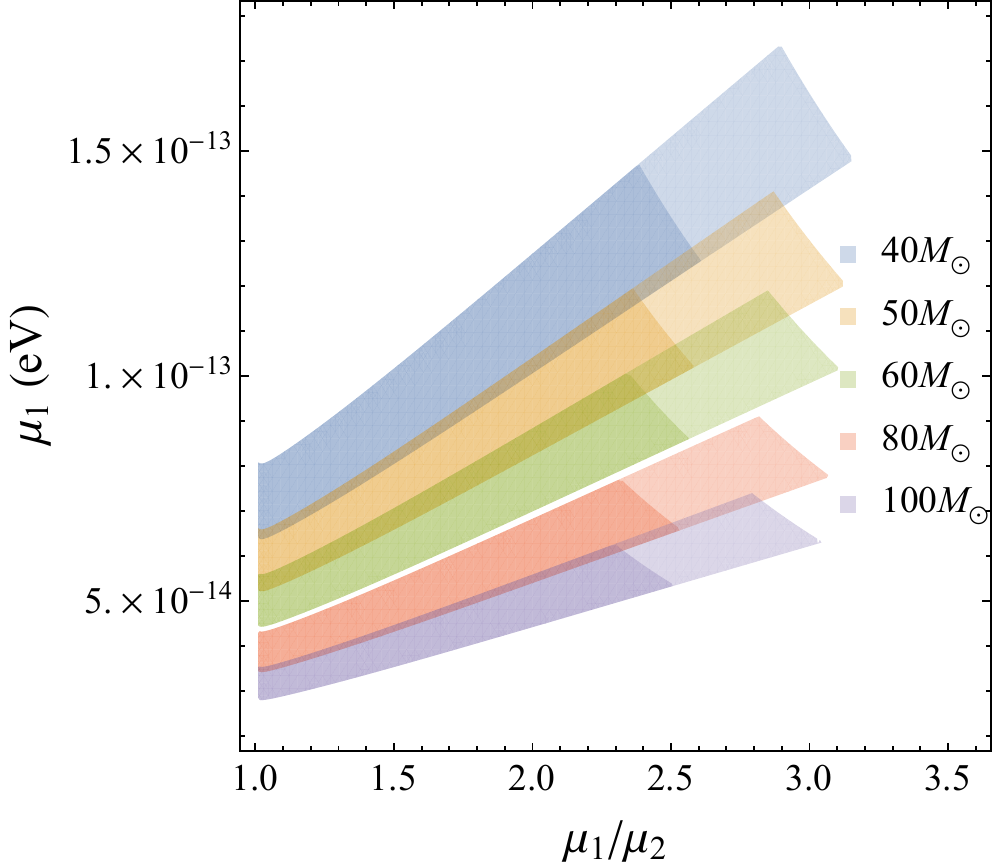}
	\caption{Values of the two axion masses for which the heavy-to-light axion cloud transition occurs between 1 and 14 Gyrs, for different values of the BH mass with $\tilde{a}_i=0.7$, and where GW emission reduces the number of heavy axions by $\leq1$ \% (darker regions) or by $\leq10$ \% (lighter regions).}
	\label{fig: racio}
\end{figure}

In Fig.~\ref{fig: racio} we show the values of the two axion masses for which the heavy-light axion transition occurs between $\sim 1$ Gyr and the present day, for different values of the BH mass in the range typical of the BH binary mergers recently detected by the LVK collaborations, assuming $\tilde{a}_i=0.7$ typical of such mergers. We also highlight the parameter regions where GW emission depletes the heavy axion cloud by less than 1\% and 10\%, roughly corresponding to mass ratios $1<\mu_1/\mu_2\lesssim 3$.



\subsection{Accretion disks}

If the BH has an accretion disk around it, this may have the opposite effect of GW emission since, following our analogy, it provides a source of ``food'' for our ecosystem of two competing axion populations. An accretion disk feeds the BH's mass and angular momentum as matter particles collide with each other and fall beyond the radius of the innermost stable circular orbit, $r_{ISCO}$, which is the minimum radius where matter can stably orbit before in-spiralling towards the horizon \cite{Bardeen:1972fi}:
\begin{equation}
	r_{ISCO}= M\bigg(3+Z_2-\sqrt{(3-Z_1)(3+Z_1+2Z_2)}\bigg),
\end{equation}
where
\begin{equation}
    \begin{aligned}
	& Z_2 = \sqrt{3\tilde{a}^2+Z_1^2}~, \\   & Z_1 = 1 +(1-\tilde{a}^2)^\frac{1}{3}((1+\tilde{a})^\frac{1}{3}+(1-\tilde{a})^\frac{1}{3}).
  \end{aligned}
\end{equation}
The BH mass increases according to (see e.g.~\cite{Barausse:2014tra}):

\begin{equation}
	\label{dmaccdt}
\dot{M}_{acc}= 2.2\times 10^{-8}f_{Edd}\Big(\frac{M(t)}{M_\odot}\Big)M_\odot \text{yr}^{-1},
\end{equation}
with $f_{Edd}$ denoting the accretion efficiency, where the Eddington limit (the maximum accretion efficiency) corresponds to $f_{Edd}=1$\footnote{This limit occurs when an equilibrium between matter accretion and radiation pressure is achieved.}. We may therefore define a timescale for accretion $\tau_{acc}=M/\dot{M}_{acc}=4.5\times10^{7}f_{Edd}^{-1}$ yrs.

Accretion increases not only the mass but also the angular momentum of the BH, with $\dot{J}_{acc}=(L/E)\dot{M}_{acc}$, where \cite{Bardeen:1970zz}
\begin{equation}
\begin{aligned}
	&L=\frac{2M}{3\sqrt{3}}\left(1+2\sqrt{3\frac{r_{ISCO}}{M}-2}\right)~,
 \\ &E=\sqrt{1-\frac{2}{3}\frac{M}{r_{ISCO}}},
\end{aligned}
\end{equation}
are the angular momentum and energy per unit mass, respectively, of a test particle at $r_{ISCO}$. This yields for the BH spin parameter in the single-field case:
\begin{equation}
	\frac{d\tilde{a}}{dt} \simeq \frac{1}{\tau_{acc}}\left(\frac{L}{E M}-2\tilde{a}\right)-\left({M_P\over M}\right)^2\Gamma N~,
\end{equation}
in the non-relativistic regime $\alpha\ll1$. Given the typically large accretion timescale, SR will initially spin down the BH until the SR condition nearly saturates, $\tilde{a}\simeq 4\alpha$. Saturation cannot, however, be completely achieved since accretion tends to spin up the BH, and the two effects eventually tend to approximately balance each other such that $d\tilde{a}/dt\simeq 0$. Given that, for low spins $L/EM\simeq 3\sqrt{3/2}\simeq 3.7$ such that the spin up due to accretion occurs at a nearly constant rate, we have that $\dot{N}=\Gamma N$ remains approximately constant, so that the particle number increases linearly in time. This quasi-stationary regime is similar to the one found in \cite{March-Russell:2022zll, Calza:2023rjt} for SR clouds around evaporating (primordial) BHs.

\begin{figure}[h!]
	\centering
	\includegraphics[width=0.48\textwidth,keepaspectratio]{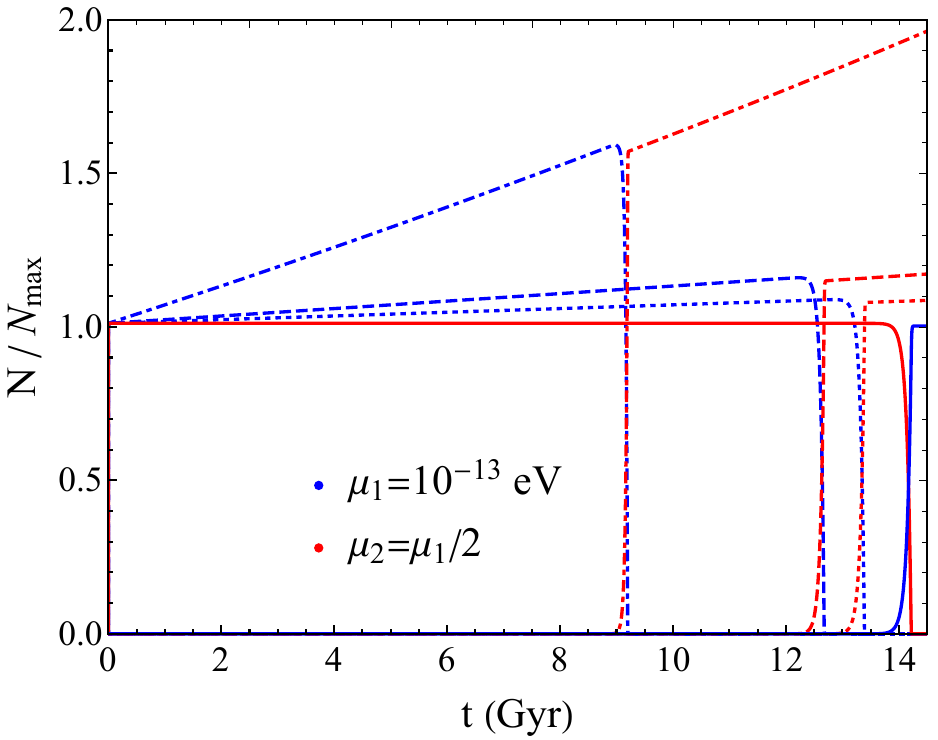}
 
	\caption{Evolution of the number of heavy (blue) and light (red) axions in  the corresponding SR clouds (normalized to their maximum capacities) for $\mu_1=10^{-13}$eV, $\mu_2=\mu_1/2$, $M=40M_\odot$ and $\tilde{a}_i=0.7$. We consider different values for the accretion efficiency $f_{Edd}=0$ (solid, no accretion), $5\times 10^{-5}$ (dotted), $10^{-4}$ (dashed) and $5\times 10^{-4}$ (dotted-dashed).} 
	\label{fig: accretionfedd}
\end{figure}

\begin{figure}[h]
\subfloat[\label{acca}]{%
 \includegraphics[width=\columnwidth]{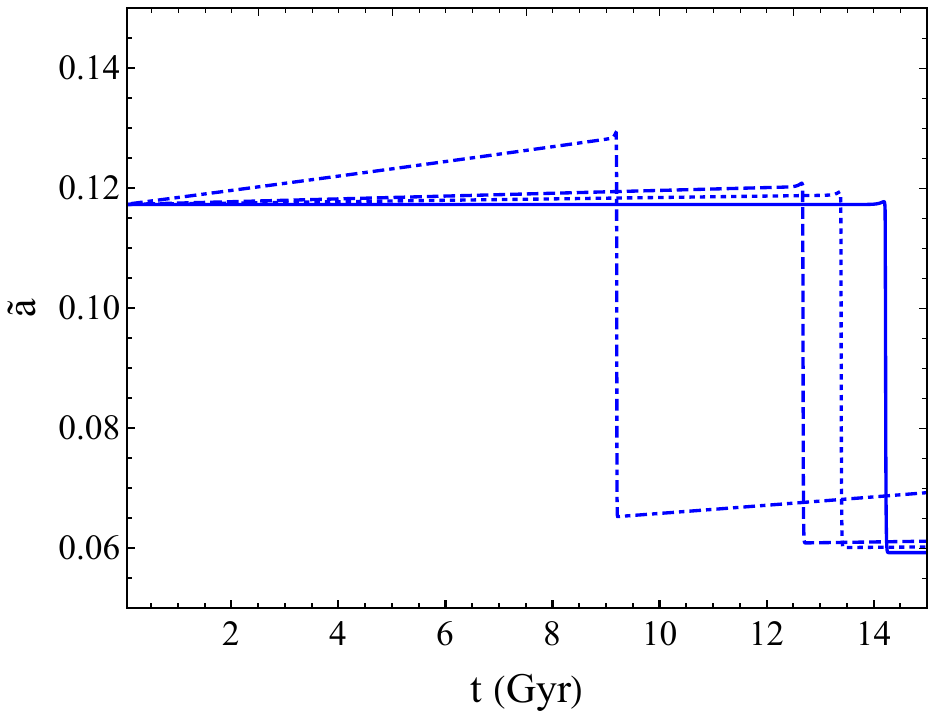}%
}\hspace*{\fill}%

\vspace{-0.5cm}
\subfloat[\label{Maccb}]{%
  \centering\includegraphics[width=0.97\columnwidth]{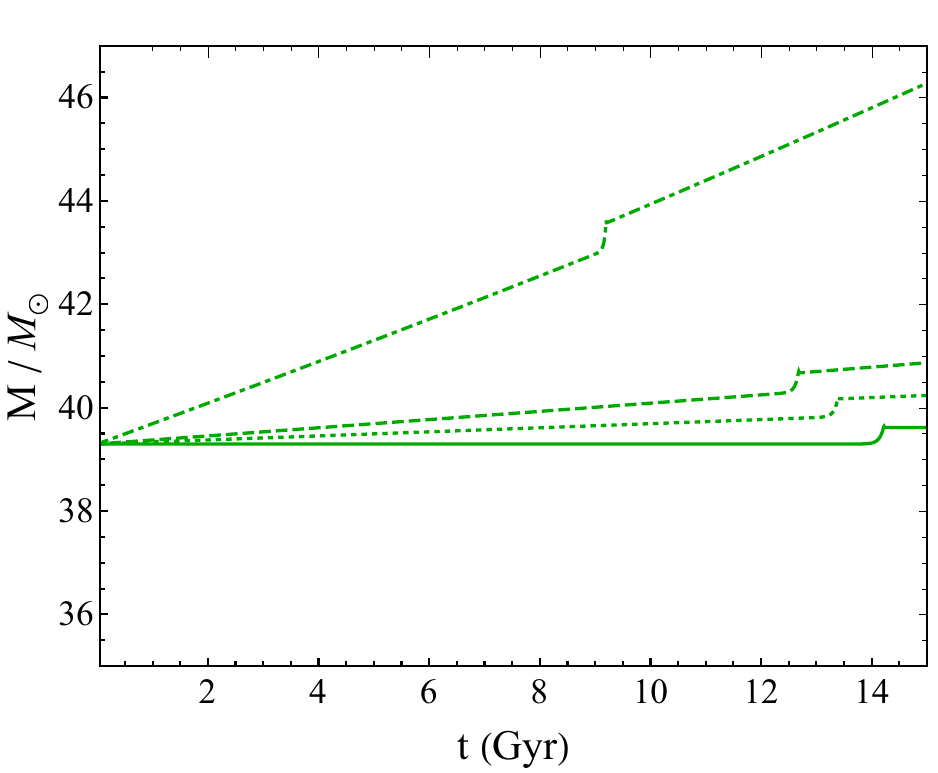}%
}
\caption{Black hole spin (a) and mass (b) evolution due to the accretion effects for the scenarios considered in Fig.~\ref{fig: accretionfedd}.}
\label{fig:7deg}
\end{figure}

In the case of two competing axion species, we expect this quasi-stationary regime to be attained for the heavier axion cloud. However, the exponential growth of the lighter axion cloud eventually destabilizes this balance, decreasing the BH spin and therefore forcing the reabsorption of the heavier cloud, as we found in the absence of significant accretion. This system nevertheless reaches a new quasi-stationary regime with a linear growth of the lighter axion cloud beyond its original maximum capacity. This is illustrated in the numerical example presented in Figs.~\ref{fig: accretionfedd} and \ref{fig:7deg} for different values of the accretion efficiency parameter (and where we have neglected GW emission to highlight the effects of accretion).

As one can see in Fig.~\ref{fig: accretionfedd}, a more efficient accretion leads to an earlier transition between the heavy and light axion SR clouds, which is mainly due to the faster increase of the BH mass and hence of the mass coupling $\alpha_2$, recalling that $\Gamma_2\propto \alpha_2^8$. In the figure we only show the dynamics up to the present time, but we note that the linear growth of the SR cloud will not last indefinitely, since accretion will eventually stop once the BH consumes the whole disk. 

We may estimate the accretion efficiency required to significantly modify the SR dynamics. The characteristic time scale for spinning up the BH with accretion is $(L/EM-2\tilde{a})^{-1}\tilde{a}\tau_{acc}\simeq 0.01\tilde{a}f_{Edd}^{-1}$ Gyrs at low BH spins. Hence, evaluating this for $\tilde{a}\simeq 4\alpha_1$ and comparing the result with the age of the Universe, we find that accretion can only significantly modify the growth of the heavy axion cloud for $f_{Edd}\gtrsim 0.004\alpha_1$. Note that most values considered in Fig.~\ref{fig: accretionfedd} are below this lower bound and that indeed the particle numbers in both heavy and light axion clouds change at most by $\mathcal{O}(1)$ factors.

An efficient accretion may potentially occur e.g.~in the case of binary stars, where one star goes supernova and collapses into a BH, while the other becomes a red-giant and eventually approaches the BH beyond the Roche limit. The latter is then broken apart by tidal effects and its remains form an accretion disk around the BH. However, in the remainder of our discussion we will focus on a population of BHs resulting from BH binary mergers, with tens of solar masses as detected by the Ligo-Virgo-Kagra collaborations, which are unlikely to be surrounded by a substantial accretion disk except perhaps in particular circumstances \cite{LIGOScientific:2016qac, Kimura:2016xmc, Mink:2017npg}. Hence, in our subsequent discussion we will neglect the effects of accretion, keeping in mind that there may be BHs for which accretion boosts the growth of the axion clouds and leads to earlier transitions between heavy and light clouds.

\section{Stochastic Gravitational Wave background}

GW detectors such as LVK in the Hz-kHz frequency range that we are interested in\footnote{A SR axion cloud with $\mu = 10^{-13}$eV produces GWs of frequency $f_{GW}=2 f \simeq 48.4$ Hz .} are sensitive to strain values $h\gtrsim 10^{-23}-10^{-22}$.  The GWs produced by SR axion clouds around stellar mass BHs produce, unfortunately, a much smaller signal, even considering BHs in our galaxy:
 \begin{equation}
	\label{eq: strain}
h\sim 6\times10^{-25}\left(\frac{\mu}{10^{-13}\ \text{eV}}\right)\left(\frac{\text{kpc}}{r}\right)\left(\frac{\alpha}{0.05}\right)^6\left(\frac{N}{10^{79}}\right)~.
\end{equation}
Nevertheless, we may expect a stochastic GW background (SGWB) resulting from all SR axion clouds accross the Universe (see e.g.~\cite{Yuan:2021ebu, Tsukada:2018mbp}). Here we will investigate whether the transition between heavy and light axion clouds may leave an imprint on the spectrum of this SGWB, and as an example we focus on SR clouds formed around BH remnants of BH binary mergers. 

The SGWB is typically characterized by the GW abundance:
\begin{equation}
\Omega_{GW}=\frac{1}{\rho_c}\frac{d\rho_{GW}}{d\log f}~,
\end{equation}
where $f$ is the GW frequency, $\rho_{GW}$ is the GW energy density and $\rho_c$ is the critical cosmological density. For isotropically distributed sources we have that
\begin{equation}
	\Omega_{GW}=\frac{f}{\rho_c}\int dM_1dM_2 dz\frac{dt_L}{dz}R(z,M_1,M_2)\frac{dE_s}{df_s}~.
\end{equation}	
Here $R(z,M_1,M_2)$ is the rate of binary mergers at a redshift $z$, and $M_1$ and $M_2$ are the progenitor BH masses. Neglecting the progenitor spin, the resulting BH remnant has a mass \cite{Berti:2007fi}:
\begin{equation}
M_{BBH} = M_1+M_2-3.62\times10^{-2}(4\nu)^2(M_1+M_2)~,
\end{equation}
with $\nu=\frac{M_1M_2}{(M_1+M_2)^2}$, and final spin \cite{Barausse:2009uz}
\begin{equation}
\tilde{a}=\nu\left(2\sqrt{3}-3.5171\nu +2.5763\nu^2\right)~.
\end{equation}
The factor $dt_L/dz$ corresponds to the ``lookback time'', that is, the time it took for the signal to arrive to us (on Earth), differentiated with respect to the redshift:
\begin{equation}
	\frac{dt_L}{dz}=\frac{1}{(1+z)H_0\sqrt{\Omega_M(1+z)^3+\Omega_\Lambda}}~,
\end{equation}
where we consider the cosmological parameters $\Omega_M=0.3$, $\Omega_\Lambda=0.7$ and $H_0=69\ \mathrm{kms^{-1}Mpc^{-1}}$ as in \cite{Yuan:2021ebu}. 
 Finally,  $dE_s/df_s$ is the energy spectrum of a single event in the source frame  with frequency $f_s$, which is related to the observed one through $f = f_s/(1+z)$. To obtain the spectrum we follow the approximation used in \cite{Yuan:2021ebu} of considering the emission of each axion cloud as a single GW event with:
\begin{equation}
		\frac{dE_s}{df_s}= E_{GW}\delta(f(1+z)-f_s)~,
\end{equation}
where $ E_{GW}=M_c^{\text{max}}-\mu N(\Delta t) = M_c^{\text{max}} \Delta t / (\Delta t +\tau_{GW}) $ is the total energy emitted by the cloud, with $M_c^{\text{max}}\simeq \mu N_{\rm{max}}$ and $\Delta t = t_0-t_{SR}$ for a single axion field, thus considering that the cloud is formed at $t_{SR}$. 

\begin{figure}[h!]
	\centering
	\includegraphics[width=0.5\textwidth,keepaspectratio]{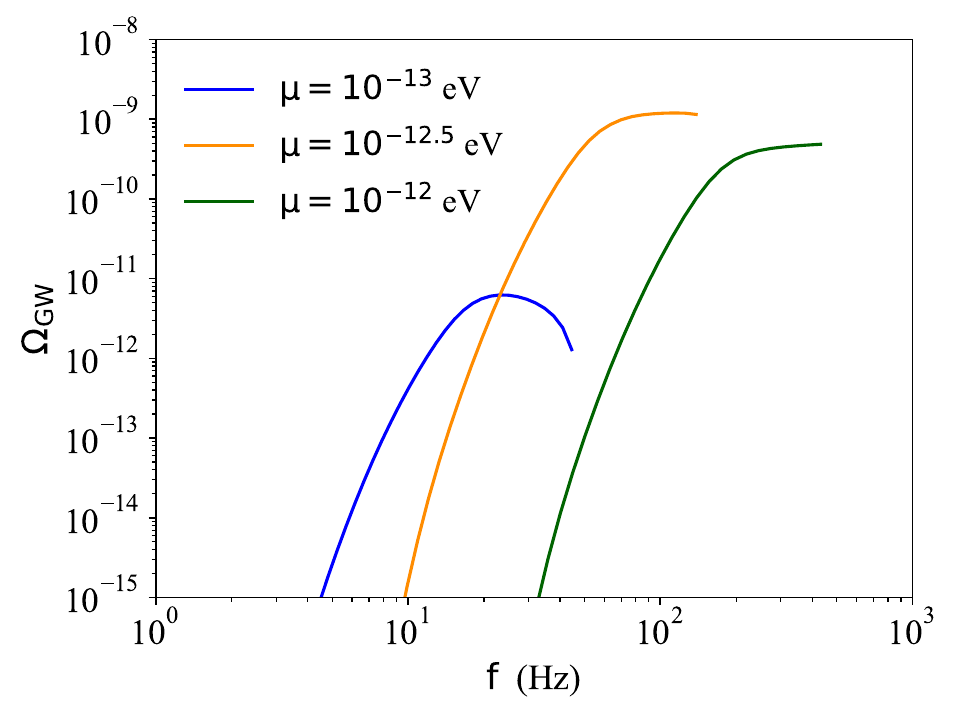}
	\caption{SGWB spectrum produced by SR axion clouds around BH-binary merger remnants, for different values of the axion mass (single axion species). }
	\label{fig: brito}
\end{figure}

The merger rate $R(z, M_1, M_2)$ at redshift $z$, corresponding to cosmic time $t$, is given by:
\begin{equation}
	\label{eq: rate}
\!R(z,M_1,M_2)\! \propto\! \int_{t_{\text{min}}}^{t_\text{max}}\!\!\frac{d\dot{n}}{dM}(z(t-t_d))P(t_d)P(M_1)P(M_2)dt_d~, 
\end{equation}
where the integration is taken from $t_{\text{min}}=50$ Myr to $t_{\text{max}}=t_0$. The factor $d\dot{n}/dM$ is the BH formation rate per unit mass that is given by \cite{Brito:2017wnc, Yuan:2021ebu, Tsukada:2018mbp}:
\begin{equation}
\frac{d\dot{n}}{dM}= \int\psi(z(t-\tau(M_*)))\phi(M_*)\delta(M_*-g^{-1}_{\text{rem}}(M,z))~,
\end{equation}
where $\phi(M_*)$ is the Salpeter initial mass function \cite{Salpeter:1955it}, which is normalized in the range $[0.1,100]M_\odot$:
\begin{equation}
	\phi(M_*)= \frac{M_*^{-2.35}}{\int_{0.1M_\odot}^{100M_\odot}M_*^{-2.35}}~,
\end{equation}
and $\tau(M_*)$ is the lifetime of the progenitor star, $M_*$, before it turns into a BH, the values of which were taken from \cite{Schaerer:2001jc}. The function $g_{\text{rem}}(M_*,z)$ gives the mass of the BH resulting from a star of mass $M_*$ given in \cite{Fryer:2011cx} and finally $\psi(z)$ is the star formation rate at redshift $z$, which can be written as:
\begin{equation}
	\psi(z)=\nu \frac{a e^{(z-z_m)(b-a)}}{(a-b)e^{-a(z-z_m)}+b}~,
\end{equation}
with $\nu=0.178M_\odot\rm yr^{-1}Mpc^{-3}$ , $a=2.37$. $b=1.80$ and $z_m=2.00$ \cite{Vangioni:2014axa}.
The next factors in Eq. ($\ref{eq: rate}$), $P(t_d)\propto 1/t_d$, $P(M_1)$ and $P(M_2)$ are the probability distributions for the time delay and BH masses, respectively. We considered the mass distributions given by the ``Broken Power Law model" of \cite{LIGOScientific:2016wof}. Finally, we normalize the merger rate at $z=0$ following the LVK local detections \cite{LIGOScientific:2016wof}:
\begin{equation}
\int R(z=0, M_1, M_2) = 23.9\; \rm yr^{-1} Gpc^{-3}~.
\end{equation} 
Given this theoretical basis we choose a range of progenitor BHs of $[3,\, 50]M_\odot$ and numerically compute the SGWB for three values of the axion mass (in the single field case), $\mu=10^{-13}\rm eV,\; 10^{-12.5}\rm eV,\; 10^{-12}\rm eV $, and the resulting spectrum is given in Fig.~\ref{fig: brito}.

Our results are in a very good agreement with those obtained in \cite{Yuan:2021ebu} neglecting the effects of higher modes, since as discussed above we are mostly interested in the parametric range where these modes do not significantly affect the dynamics of the two-axion system (see Eq. ($\ref{eq: 2srmodes}$)).

\begin{figure}[h!]
	\centering
	\includegraphics[width=0.49\textwidth]{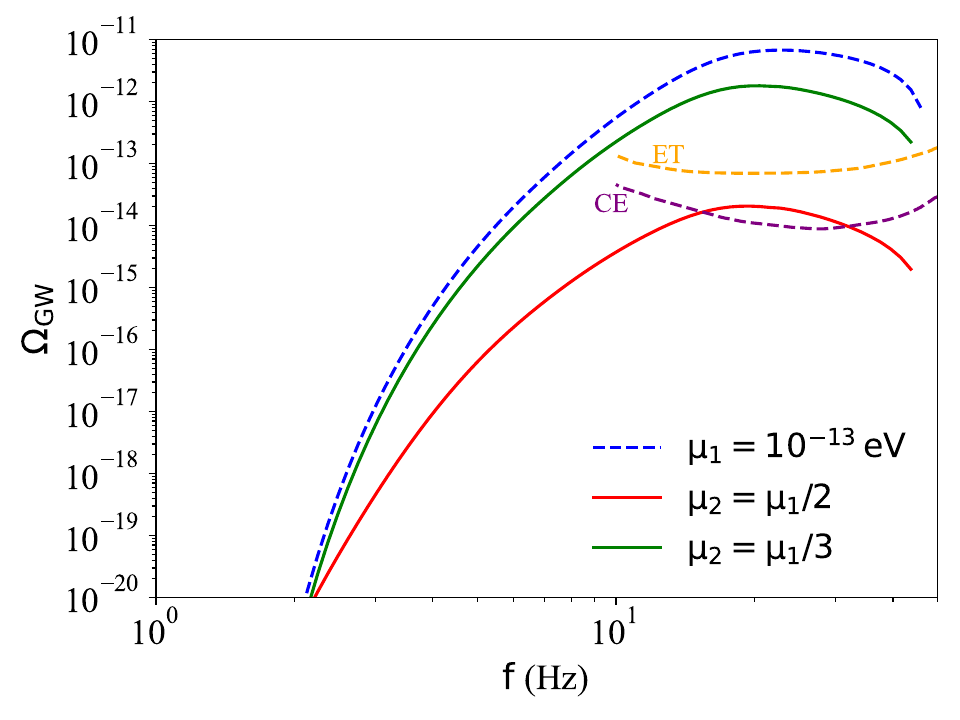}
	\caption{SGWB spectrum produced by SR axion clouds around BH-binary merger remnants, in the case of two axion species with $\mu_1=10^{-13}$ eV and $\mu_2=\mu_1/2$ (red) or $\mu_2=\mu_1/3$ (green). For comparison we also plot the SGWB spectrum for a single axion with mass $\mu_1$ (dashed-blue curve) and the expected sensitivity curves for the planned Cosmic Explorer (CE) \cite{LIGOScientific:2016wof} and Einstein Telescope (ET) \cite{Maggiore:2019uih} detectors.}
	\label{fig: twofieldssgwb}
\end{figure}

While in the single field case the cloud emits GWs until the present day, in the two-field scenario this is not the case since transitions from heavy-to-light axion SR clouds may occur. At the transition, the GW signal from the heavier axion cloud will cease and the lighter axion cloud begins to emit from that point onwards. Therefore, we considered two contributions to the source spectrum with $\Delta t = t_{\text{trans}}-t_{SR}$ for the contribution of the heavy axion clouds and $\Delta t= t_0-t_{\text{trans}}$ for the light axion clouds. We have also take into account the number of heavy axions particles that are reabsorbed by the BH and that annihilate into GWs to compute the lighter cloud mass. Our results for the SGWB spectrum in the two-axion case are shown in Fig.~\ref{fig: twofieldssgwb}.

As we can see in this figure, the main effect of including a lighter axion is to reduce the amplitude of the signal with respect to the single (heavy) axion case, but without changing the endpoint of the spectrum. This is essentially due to some of the heavy axion clouds being prematurely reabsorbed by the BHs and replaced by lighter axion ones, which therefore suppresses the signal, while other heavy axion clouds emit GWs until the present day. Recall that, for fixed axion masses, the heavy-to-light axion cloud transition occurs earlier for heavier BHs (see Fig.~\ref{fig: heatmap}). The endpoint of the spectrum is determined by the heavy axion clouds that survive until the present day, since the corresponding GWs are emitted up to $z=0$. 

\begin{figure}[h!]
	\centering
	\includegraphics[width=0.49\textwidth]{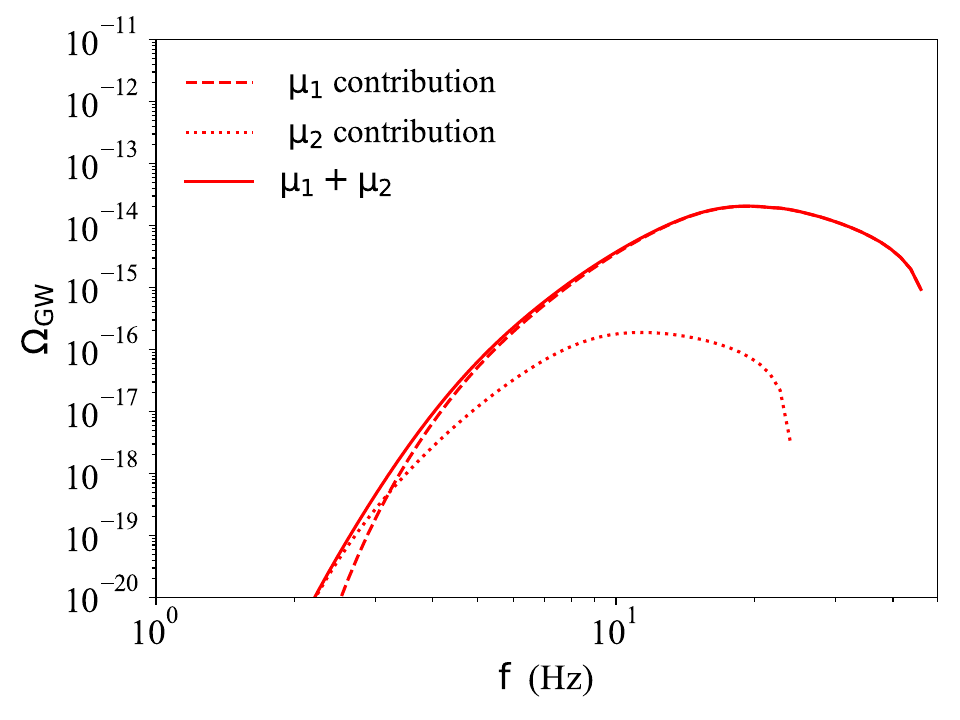}
	\caption{Contribution of the heavy (dashed) and light (dotted) axion clouds around binary-BH remnants to the total SGWB spectrum (solid) for $\mu_1=10^{-13}$ eV and $\mu_2=\mu_1/2$.}
	\label{fig: lower}
\end{figure}

We note that the contribution of the lighter axion clouds to the SGWB is typically sub-dominant, and to highlight this in Fig.~\ref{fig: lower} we plot separately the contributions of each axion species to the full SGWB spectrum, in an example where the lighter axion clouds only give a dominant contribution at low frequencies, where the signal is much harder to detect.

We also conclude that the suppression of the SGWB spectrum is more pronounced when the axion masses are closer to each other, since the heavy-to-light transition occurs earlier for larger values of $\mu_2<\mu_1$. Overall, we find that the SGWB spectrum produced by axion clouds is substantially modified with respect to the single axion scenario for mass ratios up to $\sim 3-4$.

\begin{figure}[h!]
	\centering
	\includegraphics[width=0.49\textwidth]{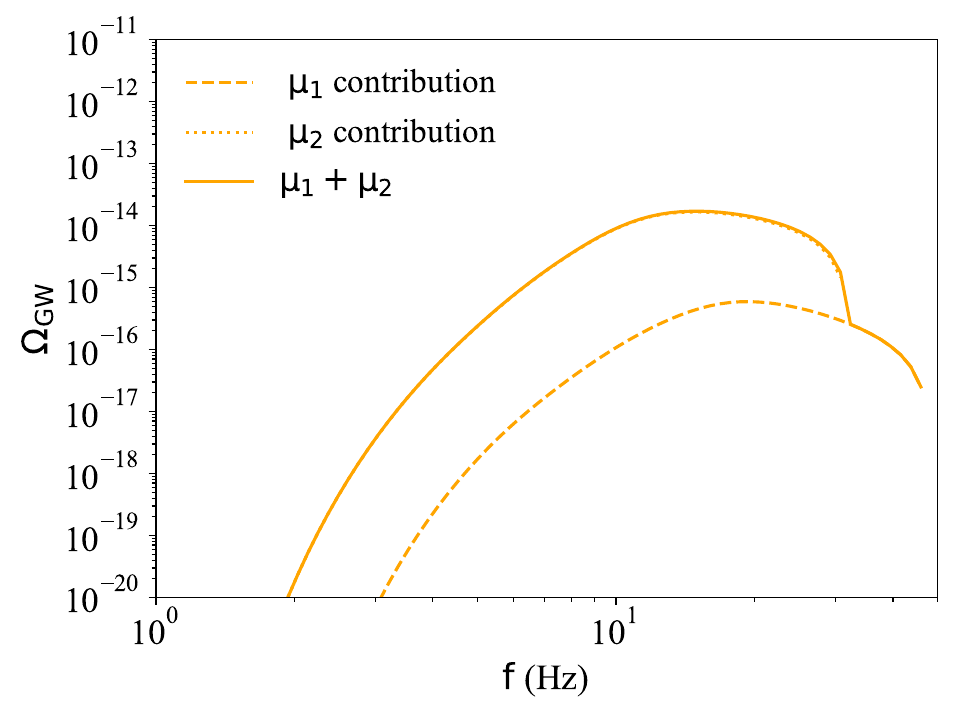}
	\caption{Contribution of the heavy (dashed) and light (dotted) axion clouds around binary-BH remnants to the total SGWB spectrum (solid) for $\mu_1=10^{-13}$ eV and $\mu_2=\mu_1/1.5$.}
	\label{fig: higher}
\end{figure}

Although it might be unlikely to find two axions with very similar masses within the axiverse, we note that the SGWB may have a unique shape for smaller mass ratios, $r<2$, since in this case there is not only a more significant suppression of the contribution of the heavy axion clouds but also a more prominent contribution of the light axion clouds, which may in fact dominate the spectrum, as illustrated in the example of Fig.~\ref{fig: higher}.

Our results therefore show that care must be taken when using the SGWB to constrain the existence of axions in a given mass range, since the spectrum may differ substantially if there is more than one axion in that mass decade. In particular, not detecting the signal expected for a given axion mass does not necessarily exclude it, since it may simply mean that there are two (or even more) axions with comparable masses. As shown in Fig.~\ref{fig: twofieldssgwb}, the planned next-generation GW detectors may be able to distinguish between single and multiple axion scenarios in the considered mass range, which may yield an interesting probe of the axiverse.

\section{Conclusion}

We have shown that a single BH may grow multiple SR axion clouds around it throughout the cosmic history if there are axion species with comparable masses, differing by a factor $< 10$. While the precise axion mass spectrum in the string axiverse depends on the details of the compact extra-dimensional geometry, one typically expects $\mathcal{O}(10^2-10^3)$ different axion species with masses spread over $<100$ orders of magnitude, so it is reasonable to expect at least some of them to lie within the same mass decade. 

These different axion clouds cannot coexist for long: heavier axion clouds form earlier but are necessarily reabsorbed by the BH once lighter axion clouds start to grow. In fact, this exhibits an interesting parallel with the dynamics of two species competing for the same resources in an ecosystem: the population of the heavier species grows faster but requires more food to survive, while the lighter species grows more slowly but needs less food to prosper. Eventually, the latter depletes food levels in the ecosystem below the minimum required for the survival of the heavier species, leading to its extinction. 

In the BH-axion case, ``food'' corresponds to BH spin and the SR condition determines whether the population of each species grows or dies away.
This parallel goes beyond a curious analogy, since the axion-BH and biological ecosystem are described by the same set of differential equations (logistic and competitive Lotka-Volterra equations in the single and multi-field cases, respectively), at least neglecting the change in the BH mass, which is typically a good approximation. We have also taken into account two effects that may influence the dynamics of realistic BHs, GW emission and accretion, with the former depleting and the latter increasing the BH spin (the amount of ``food'' in the ecosystem).

While our analysis was mostly devoted to scenarios with two axion species, our results can be easily generalized to multiple species (at least in the regime where higher excited states are not populated). The main conclusion is that there is a ``natural selection'' process in the dynamics of axion-BH SR where only the SR clouds of the lighter axion species eventually survive.

An important conclusion is that the multiple-axion dynamics may have an observational impact, as we illustrated by computing the SGWB generated by axion clouds hosted by BH binary merger remnants. Even though the contribution from the light axions is typically sub-dominant, it causes the premature reabsorption of some of the heavy axion clouds, thus suppressing the amplitude of the signal that the latter would otherwise produce, while the endpoint of the SGWB spectrum is unchanged. This must be taken into account when using GW backgrounds to probe the existence of axions in different mass ranges. In particular, if we do not 
observe the signal expected for a given axion mass, this could either mean that it does not exist or that there are multiple axions in the same mass range.

In the case of BH merger remnants, the signal suppression makes the SGWB more challenging to detect, but, at least for the axion mass range that we have focused on, it may be within the reach of planned GW observatories as the Einstein Telescope and the Cosmic Explorer.

We note that, while our discussion was mostly focused on BHs with tens of solar masses and axions in the $10^{-14}-10^{-13}$ eV mass range, our main conclusions apply to arbitrary BH and axion masses (and in fact to other types of scalar particles) as long as the SR condition is satisfied and different axion clouds have enough time to form within the cosmic history. We note that for lighter primordial BHs/heavier axions, other effects like Hawking emission or axion decays may play a relevant role in the dynamics, but we nevertheless expect the fundamental aspects of our Darwinistic view of the axiverse-BH evolution to prevail.

\vspace{0.5cm} 
\begin{acknowledgments}
D.~N.~was partially supported by a research fellowship awarded by CFisUC. This work was supported by FCT - Fundação para a Ciência e Tecnologia, I.P.~through the projects CERN/FIS-PAR/0027/2021, UIDB/04564/2020 and UIDP/04564/2020, with DOI identifiers 10.54499/CERN/FIS-PAR/0027/2021, 10.54499/UIDB/04564/2020 and 10.54499/UIDP/04564/2020, respectively.
\end{acknowledgments}

\vfill

\end{document}